\documentclass[aps,onecolumn,showpacs]{revtex4}
\usepackage[dvips]{graphics,graphicx,epsfig}
\usepackage{doublespace}
\usepackage{amsfonts}

\begin{document}

\def\beqn{\begin{equation}}
\def\eeqn{\end{equation}}
\def\beqnar{\begin{eqnarray}}
\def\eeqnar{\end{eqnarray}}
\newcommand{\ket}[1]{$\vert${#1}$\rangle$}
\newcommand{\mket}[1]{\vert{#1}\rangle}
\newcommand{\mbra}[1]{\langle{#1}\vert}
\newcommand{\tfrac}[2]{{\textstyle\frac{#1}{#2}}}
\newcommand{\ignore}[1]{}
\def\up{|\uparrow\,\rangle}
\def\dn{|\downarrow\,\rangle}
\def\upd{\langle\, \uparrow |}
\def\dnd{\langle\, \downarrow |}
\newcommand{\mb}[1]{\mbox{\boldmath{$#1$}}}
\def\ba{\begin{array}}
\def\ea{\end{array}}
\newcommand{\wdg}{\! \wedge \!}
\newcommand{\crs}{\! \times \!}
\newcommand{\scp}{\! \ast \!}
\newcommand{\dt}{\! \cdot \!}
\newcommand{\etal}{{\em et al. }}
\newcommand{\eqn}[1]{(\ref{#1})}

\title{Experimental Demonstration of Quantum Lattice Gas Computation}

\author{Marco A. Pravia}
\affiliation{Department of Nuclear Engineering, Massachusetts Institute of Technology, Cambridge, MA  02139}
\author{Zhiying Chen}
\affiliation{Department of Nuclear Engineering, Massachusetts Institute of Technology, Cambridge, MA  02139}
\author{Jeffrey Yepez}
\affiliation{Air Force Research Laboratory, Hanscom Field, MA  01731}
\author{David G. Cory\footnote{To whom correspondence should be addressed.
         E-mail: dcory@mit.edu }}
\affiliation{Department of Nuclear Engineering, Massachusetts Institute of Technology, Cambridge, MA  02139}

\begin{abstract}

We report an ensemble nuclear magnetic resonance (NMR) implementation
of a quantum lattice gas algorithm for the diffusion equation.
The algorithm employs an array of quantum information processors 
sharing classical information, a novel architecture referred 
to as a type-II quantum computer.  This concrete implementation 
provides a test example from which to probe the strengths and 
limitations of this new computation paradigm.  The NMR
experiment consists of encoding a mass density onto an array of 16 two-qubit 
quantum information processors and then following the computation through 7 
time steps of the algorithm.  The results show good agreement with the analytic 
solution for diffusive dynamics.  We also describe numerical simulations
of the NMR implementation.  The simulations aid in determining
sources of experimental errors, and they help define the limits
of the implementation. 

\end{abstract}


\maketitle

\section{Introduction}

The advent of fast quantum algorithms\cite{GenQCRefs} has spawned a broad 
search for new algorithms that utilize the novel features of quantum 
information. Among the new proposals are quantum lattice gas (QLG) algorithms, 
which, in analogy to their classical counterparts, make use of arrays of 
interacting sites to perform useful calculations.  In the quantum case, 
however, the sites behave quantum mechanically, while the site-to-site 
interactions can be either classical\cite{YepezLatticegas} or 
quantum mechanical\cite{Meyer_QLGA}.  New algorithms have been devised 
to solve selected computational problems such as the diffusion 
equation\cite{YepezDiffusion, Meyer-Diff} and the Schr\"{o}dinger 
equation\cite{QLG-Schrodinger}.  In the case where the quantum 
mechanical sites (or nodes) communicate with each other classically,
the required architecture for QLG algorithms has been termed a 
type-II quantum computer\cite{YepezTypeII}.  

A type-II device is essentially a classically parallel computer, with the 
exception that the computing elements follow the rules of quantum 
mechanics.  The advantage gained from the classical network is 
completely analogous to the improvement gained in a classical, 
massively-parallel architecture.  However, the use of quantum 
mechanical nodes introduces several notable differences.  Classical
lattice gas algorithms become unstable (and unusable) when the relevant 
transport coefficient is reduced or when nonlinearities are increased.  
In the quantum case, the transport coefficient and degree of nonlinearity
can be varied at will using the appropriate quantum operations at 
each site.  In addition, the quantum algorithms typically require
a smaller number of qubits per site than do the classical algorithms.  
Finally, the family of QLG algorithms can handle increasingly complex 
calculations as the number of (qu)bits per site is increased.  For 
example, when two qubits are present in a site, the QLG algorithms can 
solve the relatively simple diffusion equation in one, two, or three 
dimensions \cite{YepezDiffusion} and the more difficult nonlinear 
Burgers equation in one dimension \cite{YepezBurgers}.  With four 
qubits per site, the QLG algorithms can solve coupled nonlinear field 
equations governing the velocity and magnetic fields of one-dimensional 
magnetohydrodynamic turbulence \cite{YepezMHD}.    With 
six qubits per site, the QLG algorithms can model the nonlinear 
Navier-Stokes equations in two dimensions governing a viscous fluid 
\cite{yepez-lncs99}.  A more complete description of type-II quantum 
computers and their scaling properties has been given by Jeffrey Yepez
\cite{YepezBurgers}\cite{YepezMHD}.

Here, we present a methodology for implementing a quantum lattice gas
algorithm on a nuclear magnetic resonance (NMR)\cite{NMR} type-II architecture.
In this implementation, we encode a discrete mass density onto 
distinct spatial locations of a liquid-state sample.  
We use magnetic field gradients to discriminate between 
locations in the sample, thus creating an array of addressable
ensemble NMR quantum information processors.
In addition, we use radio frequency (RF) pulses and methods
learned from previous work\cite{NMRQIPBasics1, NMRQIPBasics2} to execute
the necessary operations in each quantum processor.  
The result is a concrete implementation examining the necessary 
control for realizing a quantum lattice-gas algorithm using NMR techniques.

\section{Lattice Gas Algorithms}
The lattice gas method is a tool of computational physics used to model 
hydrodynamical flows that are too large for a standard low-level 
molecular dynamics treatment and that contain discontinuous interfacial 
boundaries that prevent a high-level partial differential equations 
description\cite{LGOrigins1, LGOrigins2, LGOrigins3, LGOrigins4}.  
The basic idea underlying the lattice gas method is to 
statistically represent a macroscopic scale time-dependent 
field quantities by ``averaging'' over repeated instances of a system of 
artificial microscopic particles scattering and propagating throughout a 
lattice of interconnected sites.  A particular instance of the system has 
many particles distributed over the lattice sites.  Multiple particles 
may coexist at each site at a given time, and each particle carries a 
unit mass and a unit momentum of energy.  Particles interact on site 
via an artificial collision rule which exactly conserves the total 
mass, momentum, and energy at that site.  The movement of particles 
along the lattice is prescribed by a streaming operation that 
shifts particles to nearest neighboring sites,
thus endowing the particles with the property of momentum.  The 
lattice gas algorithm encapsulates the microscopic 
scale kinematics of the particles scattering on site and moving 
along the lattice.  The mean-free path length between collisions 
is about one lattice cell size and the mean-free time between 
collision elapses after a single update.  This is computationally 
simple in comparison to molecular dynamics where many thousands 
of updates are required to capture such particle interactions.

The mesoscopic evolution is obtained by taking the ensemble 
average over many instances of microscopic realizations.  At the 
mesoscopic scale, the average presence of each particle type is defined
by an occupation probability.  In addition, the 
microscopic collision and streaming rules translate into the 
language of kinetic theory\cite{KinTheory1,KinTheory2,KinTheory3}.  
The behavior of the system is described 
by a transport equation for the occupation probabilities, and this 
equation is a discrete Boltzmann equation called the 
{\it lattice Boltzmann equation} \cite{LatBoltz1,LatBoltz2,LatBoltz3}.  

The lattice Boltzmann equation further translates into a macroscopic,
continuous, effective field theory by letting the cell size approach
zero (the limit of infinite lattice resolution called the {\it continuum limit}).
At the macroscopic scale, partial differential equations describe the
evolution of the field, admitting solutions such as propagating 
sound wave modes and diffusive modes.  The passage 
of the Boltzmann equation to the effective field 
theory begins by expanding the occupation probabilities, which 
have a well-defined statistical functional form, in terms of the 
continuous macroscopic variables, such as the mass density $\rho$ 
(and the velocity or energy field if they are defined in the model).  
This expansion usually is carried out perturbatively in a small 
parameter such as the Knudsen number (ratio of mean-free path to the 
largest characteristic length scale) or the Mach number (ratio of 
the sound speed to the largest characteristic flow speed) in a 
fashion analogous to the Chapman-Enskog expansion of kinetic theory \cite{ChapEn1,ChapEn2,ChapEn3}.  
Conversely, and self-consistently, the macroscopic field quantities 
can also be expressed as a function of the mesoscopic occupation 
probabilities--for example, the mass density at some point is a 
sum over the occupation probabilities in that vicinity.

Quantum lattice gas algorithms are generalizations of the classical 
lattice gas algorithms described above where quantum bits are used 
to encode the occupation probabilities and where the principle of 
quantum mechanical superposition is added to the artificial 
microscopic world.   In this quantum case, the mesoscopic occupation 
probabilities are mapped onto the wave functions of quantum mechanical 
sites.  In the case where the quantum lattice gas describes a 
hydrodynamic system when the time evolution of the flow field 
is required, we must periodically measure these 
occupation probabilities and the quantum lattice gas algorithm 
becomes suited to a type-II implementation.    Such type-II algorithms 
have been shown to solve dynamical equations such as the 
diffusion equation \cite{YepezDiffusion}, the Burgers 
equation \cite{YepezBurgers}, and magnetohydrodynamic Burgers 
turbulence \cite{YepezMHD}.  As a first exploration of a type-II 
architecture using NMR, we implemented a QLG model of diffusive 
dynamics in one dimension.

\section{Solving the 1-D Diffusion Equation}

The quantum lattice gas algorithm that solves the 1-D diffusion equation 
derives from a classical lattice gas of particles moving up and down
a 1-D lattice\cite{YepezDiffusion}.  The motion of the particles occurs in discrete steps 
(streaming phase), and the particles have a probability of changing directions (collision).  
When the collisions are such that the particles reverse directions half of the time,
then the continuum effective field theory that emerges obeys diffusive dynamics.  In 
this case, the motion of an individual particle is a random walk, and an arbitrary 
initial distribution of particles will diffuse isotropically as a function of time.

The lattice gas described above is summarized by the Boltzmann equation
\begin{equation}
\label{BoltzmannEq}
f_{1,2}(z\pm\Delta z, t+\Delta t) = f_{1,2}(z,t) + \Omega_{1,2}(z,t),
\end{equation}
where the left-hand side denotes the occupation of the lattice as a function of the 
previous lattice configuration and where the collision term is 
\beqn
\Omega_{1,2} = \pm\frac{1}{2} \left[ f_1\left(1-f_2\right) -  f_2\left(1-f_1\right) \right]
\eeqn
The variables $f_1 \equiv f_1(z,t)$ and $f_2 \equiv f_2(z,t)$ are the occupation 
probabilities for finding upward- and downward-moving particles, respectively, 
at the site location $z$ and time $t$.  The time step is denoted by $\Delta t$, 
while the lattice spacing is given by $\Delta z$.  The collision term changes 
the direction of some particles, and it is responsible for the diffusive behavior.

The interesting macroscopic quantity of the lattice gas is the mass density 
field, $\rho$, defined as the sum of upward- and downward-moving particles
\beqn
\rho(z, t) = f_1(z,t) + f_2(z,t).
\eeqn
The ambiguity in assigning the mass density between the two occupation 
probabilities is resolved by a constraint for local equilibrium demanding
that the mass density be initially distributed equally 
\begin{equation} \label{localEquil}
f^{\hbox{\tiny eq}}_1(n,0) = f^{\hbox{\tiny eq}}_2(n,0) = \frac{1}{2} \rho (n\Delta z, 0).
\end{equation}
After a single time step, the occupation probabilities $f_1$ and $f_2$ evolve 
according to (\ref{BoltzmannEq}), resulting in a new mass density 
\beqn
\rho(z,t+\Delta t) = \frac{1}{2}\left[\rho(z+\Delta z,t)+\rho(z-\Delta z,t) \right].
\eeqn
The first finite-difference in time of the mass density field
is then written as
\beqn \label{finiteDiff}
\rho(z,t+\Delta t)-\rho(z,t)=\frac{1}{2}\left[\rho(z+\Delta z,t)-2\rho(z,t)+\rho(z-\Delta z,t)\right].
\eeqn
In the limit where the lattice cell size and the time step approach zero 
($\Delta z\rightarrow 0$ and $\Delta t\rightarrow 0$), the mass density field
becomes continuous and differentiable.  The second-order Taylor expansion of equation 
(\ref{finiteDiff}) about $z$ and $t$ can thus be written in the differential form
\beqn
\frac{\partial \rho(z,t)}{\partial t}=\left(\frac{\Delta z^2}{2 \Delta t}\right) \frac{\partial^2 \rho(z,t)}{\partial z^2}
\eeqn
where it is now evident that $\rho$ evolves according to the diffusion 
equation with a constant transport coefficient $\frac{\Delta z^2}{2 \Delta t}$.  

Finally, in this implementation we consider an initial mass density ${\rho(z,t=0)}$
whose evolution obeys the periodic boundary condition $\rho(z,t) = \rho(z+L,t)$,
where $L$ is the length of the lattice.  As a result, the initial mass density
diffuses until the total mass is evenly dispersed throughout the lattice.

The corresponding quantum lattice gas algorithm description begins by encoding the occupation probabilities, and thus the mass density, in the states of a lattice of quantum objects.  The streaming and collision operations are then a combination of classical and quantum operations, including measurements.  The aim of the algorithm is to take an initial mass density field and to evolve its underlying occupation probabilities according to the Boltzmann equation (\ref{BoltzmannEq}).  A schematic of the entire quantum algorithm is shown in Fig. 1.  
A single time step of the algorithm is decomposed into four sequential operations:
\begin{enumerate}
\item encoding of the mass density
\item application of the collision operator $\hat{C}$ at all sites
\item measurement of the occupation numbers
\item streaming to neighboring sites.
\end{enumerate}
These operations are repeated until the mass density field has evolved for the desired number of time steps.  In the first time step, the encoding operation specifies the initial mass density profile, while in all the subsequent steps the encoding writes the results of the previous streaming operation.  The final time step ends with the readout of the desired result, so operation 4 is not performed.

Each occupation probability is represented as the quantum mechanical
expectation value of finding a two-level system, or qubit, in its 
excited state $\mket{1}$.  As a result, the state of the qubit 
encoding the value $f_a(z,t)$ is
\beqn
\mket{f_a(z,t)} = \sqrt{f_a(z,t)} \mket{1} + \sqrt{1-f_a(z,t)} \mket{0}.
\eeqn
It follows that a single value of the mass density is recorded in two
qubits, one for each occupation number.  The combined two-qubit
wave function for a single node becomes
\beqnar
|\psi(z,t)\rangle&=&\sqrt{f_1f_2}|11\rangle + \sqrt{f_1(1-f_2)}|10\rangle+ \\
        & &\sqrt{(1-f_1)f_2}|01\rangle + \sqrt{(1-f_1)(1-f_2)}|00\rangle \nonumber.
\eeqnar
The kets $|00\rangle$, $|01\rangle$, $|10\rangle$, and $|11\rangle$ span 
the joint Hilbert space of the two qubits, and this is the largest 
dimension space over which quantum superpositions are allowed.  
As with the classical algorithm, the constraint for local equilibrium
(\ref{localEquil}) forces the initial occupation probabilities 
at a node to be half of the corresponding mass density value.

The occupation numbers encoded in the two-qubit wave function 
$|\psi(z,t)\rangle$ can be recovered by measuring the expectation
value of the number operator $\hat{n}_a$, as given in
\begin{equation} \label{OccupationNumbers}
f_a(z, t) = \langle\psi(z,t)|{\hat{n}_a}|\psi(z,t)\rangle ,
\end{equation}
where $\hat n_1=\hat n \otimes{\bf 1}$, $\hat n_2={\bf 1}\otimes \hat n$, 
where ${\bf 1}$ is the $2\times 2$ identity matrix, and where the 
action of the single-qubit number operator $\hat n$  returns $1$ if 
qubit is in its excited state and $0$ for the ground state.  

The encoded occupation probabilities evolve as specified by the Boltzmann equation by the combined action of the collision operator, the measurement, and streaming.  The collision operator contributes by taking the local average of the two occupation probabilities.  This averaging (not to be confused with statistical coarse-grain averaging, time averaging, or ensemble averaging) is done by choosing the the collision operator $\hat{C}$ to be the ``square-root of swap'' gate, written as
\begin{equation}
\hat{C} = \left(
\begin{array}{cccc} 
        1 & 0 & 0 & 0 \\      
       0 & \frac{1}{2}+\frac{i}{2} & \frac{1}{2}-\frac{i}{2}& 0\\
       0 & \frac{1}{2}-\frac{i}{2} &  \frac{1}{2}+\frac{i}{2}& 0\\
       0 & 0 & 0 & 1  
\end{array} \right)
\end{equation}
in the standard basis.  The same collision is applied simultaneously at every site, 
resulting in 
\begin{equation}
|\psi'(z,t)\rangle=\hat{C} |\psi(z,t)\rangle . 
\end{equation}
Using (\ref{OccupationNumbers}), the intermediate
occupation probabilities of the wave function $|\psi'(z,t)\rangle$ are
\begin{equation} 
f_a'(z, t) = \frac{1}{2} \left( f_1 + f_2 \right)
\end{equation}
as required for $a=1,2$.  The third operation physically measures these intermediate occupation probabilities $f_a'(z,t)$ at all the sites.   
If the algorithm is performed on individual quantum systems, then the values are obtained by averaging over many strong quantum measurements of identical instances of each step.  However, when the algorithm is performed using a sufficiently large ensemble of quantum systems, as in the case of NMR, then a single weak measurement of the entire ensemble can provide sufficient precision to obtain $f_a'(z,t)$.  
A single time step is completed with the streaming of the occupation probabilities to the nearest neighbors, according to the rule
\begin{eqnarray}
\label{streamOcc1} f_1(z-\Delta z,t+\Delta t)=f_1'(z,t)  \\
\label{streamOcc2} f_2(z+\Delta z,t+\Delta t)=f_2'(z,t) 
\end{eqnarray}
The information on the two qubits is shifted to the neighboring sites in opposite directions.  The streaming operation is a classical step causing global data shifting, and it is carried out in a classical computer interfaced to the quantum processors.  Together, the last three operations result in 
\begin{eqnarray}
f_{1,2}(z\pm\Delta z,t+\Delta t)= \frac{1}{2} \left[ f_1(z,t) + f_2(z,t) \right]  \nonumber \\
\end{eqnarray}
which is the exact dynamics described by the Boltzmann equation (\ref{BoltzmannEq}).

\section{NMR Implementation}

\subsection{Spin System and Control}
The goal of the NMR implementation is to experimentally explore the 
steps outlined by the diffusion QLG algorithm.  For this two-qubit problem,
we chose a room-temperature solution of isotopically-labeled chloroform
($^{13}$CHCl$_3$), where the hydrogen nucleus and the labeled carbon nucleus
served as qubits 1 and 2, respectively\cite{ChloroRef}.  The chloroform sample was divided
into 16 classically-connected sites of two qubits each, creating an 
accessible Hilbert space larger than would be available with 32 
non-interacting qubits.

The internal Hamiltonian of this system in a strong and homogeneous 
magnetic field $B_0$ is
\beqn  \label{systemHamiltonian}
H_{internal} = -\frac{1}{2} \left( \gamma_H B_0 \right) \sigma^1_z - \frac{1}{2} \left( \gamma_C B_0 \right) \sigma^2_z + \frac{\pi J}{2} \sigma^1_z \sigma^2_z
\eeqn
where the first two terms represent the Zeeman couplings of the spins with 
$B_0$ and the last term is the scalar coupling between the two spins.  
The operators of the form $\sigma^{a}_{k}$ are Pauli spin operators
for the spin $a$ and the Cartesian direction $k$.  The choice of
chloroform is particularly convenient because the different 
gyromagnetic ratios, $\gamma_H$ and $\gamma_C$, generate 
widely spaced resonant frequencies.  As a result, a RF pulse 
applied on resonance with one of the spins
does not rotate, to a very good approximation, the other spin.  In the 7 T 
magnet utilized for the implementation, the hydrogen and carbon 
frequencies were about 300 MHz and 75 MHz, respectively.  The widely
spaced frequencies allow us to write the two RF control Hamiltonians 
as acting on the two spins independently.  The 
externally-controlled RF Hamiltonians are written as
\begin{eqnarray} \label{RFHamiltonian}
H_{RF}^{a}(t) &=& -\frac{1}{2} \left[w^a_x(t) \sigma^a_x +w^a_y(t) \sigma^a_y\right] . 
\end{eqnarray}
The RF Hamiltonians generate arbitrary single-spin rotations with high 
fidelity when the total nutation frequencies
\beqn
\nu^a_{RF} = \frac{1}{2 \pi} \sqrt{[w^a_x]^2 +[w^a_y]^2 }
\eeqn
are much stronger than $J$, the scalar coupling constant.  The scalar
coupling Hamiltonian and the single-spin rotations permit the implementation
of a universal set of gates, and they are the building blocks for constructing
more involved gates such as the collision operator $\hat{C}$\cite{NMRQIPBasics1, NMRQIPBasics2}.

The lattice of quantum information processors is realized by superimposing a 
linear magnetic field gradient on the main field $B_0$, adding a position
dependent term to the Hamiltonian having the form
\begin{equation}
H_{gradient}(z) = -\frac{1}{2} \left( \gamma_H  \frac{\partial B_z}{\partial z} z \right) \sigma^1_z- \\
\frac{1}{2} \left( \gamma_C \frac{\partial B_z}{\partial z} z \right) \sigma^2_z .
\end{equation}
The variable $z$ denotes the spatial location along
the direction of the main field, while the constant $\frac{\partial B_z}{\partial z}$
specifies the strength of the gradient.  The usefulness of this Hamiltonian 
can be appreciated by noticing that the offset frequencies 
$\Delta \Omega_{H,C} =  \gamma_{H,C} \left( \frac{\partial B_z}{\partial z} \right) z$
of the spins vary with position when the gradient field is applied. 
Spins at distinct locations can thus be addressed
with RF fields oscillating at the corresponding frequencies.  In 
this way, the magnetic field gradient allows the
entire spin ensemble to be sliced into a lattice of smaller, individually
addressable sub-ensembles.

Using the coupling, RF, and gradient Hamiltonians described above, together 
with the appropriate measurement and processing tools, we can now
describe in detail how the four steps of the diffusion QLG algorithm translate to 
experimental tasks.  The lattice initialization step (1) uses the magnetic
field gradients to establish sub-ensembles of varying resonant frequency
addressable with the RF Hamiltonians.  The collision step (2) makes use of 
both the RF and the internal coupling Hamiltonians to generate the desired
unitary operation $\hat{C}$\cite{NMRQIPBasics1, NMRQIPBasics2}.  The 
readout (3) is accomplished by measuring the spins in the presence of 
a magnetic field gradient.  And finally, the streaming operation (4) is
performed as a processing step in a classical 
computer in conjunction with the next initialization step.  

\subsection{Lattice Initialization}

The initialization of the lattice begins by transforming the equilibrium 
state of the ensemble into a starting state amenable for quantum
computation.  At thermal equilibrium, the density matrix is
\beqn
\sigma_{thermal} = \text{exp}\left[-\frac{H_{internal}}{k_B T}\right] \approx \frac{\mathbf{1}}{2^2} + \epsilon  \left[ \frac{\gamma_H}{\gamma_C} \sigma^1_z + \sigma^2_z \right]
\eeqn
where $\epsilon$ has a value on the order of $10^{-5}$.  The equilibrium
state is highly mixed and the two spins have unequal magnetizations.  To 
perform quantum computations, it is convenient to transform the equilibrium 
state into a pseudo-pure state \cite{CoryPP,IkePP}, a mixed state whose deviation 
part transforms identically to the corresponding pure state and, when measured,
returns expectation values proportional to those that would be obtained 
by measuring the underlying pure state.  Two transformations create the
starting pseudo-pure state $|00\rangle$ from the thermal state.  First, the
magnetizations of the two spins are equalized,
\beqn
\sigma_{thermal} \stackrel{Equalize}{\longrightarrow} \sigma_{equal} = \\
\frac{\mathbf{1}}{2^2} + \frac{\epsilon}{2} \left( 1+ \frac{\gamma_H}{\gamma_C} \right) \left[ \sigma^1_z + \sigma^2_z \right]
\eeqn
followed by a pseudo-pure state creation sequence that results in 
\beqn
\sigma_{equal} \stackrel{Pseudo-pure}{\longrightarrow} \sigma_{pp} = \\
\frac{\mathbf{1}}{2^2} + \epsilon \frac{\sqrt{3}}{4\sqrt{2}} \left(  1+ \frac{\gamma_H}{\gamma_C}   \right) \\
 \left[ \sigma^1_z + \sigma^2_z + \sigma^1_z \sigma^2_z \right] .
\eeqn
The equalization and pseudo-pure state creation sequences are described
in detail in reference \cite{PPEqual}.  For clarity, we define the constant
in front of the brackets to be $\epsilon'$, allowing us to write the 
pseudopure state $\sigma_{pp}$ in terms of the desired spinor 
$|00\rangle$ as
\beqn
\sigma_{pp} = \left( \frac{1}{4}- \epsilon' \right) {\mathbf 1} + \epsilon' |00\rangle \langle00| .
\eeqn
Expressed in this manner, it is now seen that a unitary transformation
applied to $\sigma_{pp}$ acts trivially on the term proportional to the 
identity, but it evolves the term $|00\rangle\langle00|$ as it would 
a pure state.

Individually addressing the sites of the lattice, as depicted in Fig. 1, is 
accomplished by selectively addressing adjacent slices of the cylindrical sample.  
The procedure is related to slice-selection in magnetic resonance imaging 
(MRI)\cite{MRSliceSelection}, and it works by applying the gradient Hamiltonian
in the presence of suitably shaped RF pulses.  First, consider the
Hamiltonian for a one-spin system subjected to a linear magnetic 
field gradient in the $z$-direction and to a time-dependent RF pulse 
applied in the $y$-direction.  In this case, the Hamiltonian is 
\beqn
H_{RF,G}(z,t) = -\frac{1}{2} \left( \gamma \frac{\partial B_z}{\partial z} z \right) \sigma_z \\
-\frac{1}{2} w_y(t) \sigma_y
\eeqn
where the $\sigma_z$ term is the linearly-varying static field and 
the $\sigma_y$ term is the time-dependent RF.  The Hamiltonian 
$H_{RF,G}(z,t)$ does not commute with itself at all times, 
so a closed-form and exact solution cannot be easily given without
specifying the function $w_y(t)$.  A valuable approach, however, is to consider
the approximate evolution generated by $H_{RF,G}(z,t)$ during infinitesimal 
periods of the RF pulse.  To first order, the evolution during the initial period 
$\Delta t$ becomes
\beqn
U_{RF,G}(z,t=\Delta t) \approx \text{exp} \left[i \frac{1}{2} \left(\gamma \frac{\partial B_z}{\partial z} \Delta t \right) z ~ \sigma_z \right] \text{exp} \left[i \frac{w_y(\Delta t) \Delta t}{2}  \sigma_y \right]
\eeqn
By defining the term in the parenthesis as $\Delta k_z \equiv \gamma \frac{\partial B_z}{\partial z} \Delta t$,
the evolution of an initial density matrix $\sigma_z$ through a single period becomes
\beqn
U_{RF,G} \sigma_z U_{RF,G}^\dagger \approx \text{exp} \left[ i \frac{\Delta k_z z}{2} \sigma_z \right] \sigma_x \text{exp} \left[-i \frac{\Delta k_z z}{2} \sigma_z \right] w_y(\Delta t) \Delta t +  \sigma_z 
\eeqn
where small angle approximations have been made.  The first term is a spatial helix of 
the $x$ and $y$ magnetizations having a wavenumber $\Delta k_z$.  The second term
is the first order approximation to the magnetization remaining in the state $\sigma_z$.  
Another period of evolution will affect the $\sigma_z$ term as described, creating 
a new magnetization helix with wavenumber $\Delta k_z$.  In addition, the initial helix 
will have its wavenumber increased by an amount $\Delta k_z$.  The final result 
over many periods is the formation a shaped magnetization profile having 
many components
\beqn \label{KSpaceEq}
\sigma_z \rightarrow \sum_{n=1}^{N}  \text{exp} \left[i \frac{n \Delta k_z z}{2}  \sigma_z \right]  \sigma_x  \text{exp} \left[-i \frac{n \Delta k_z z}{2}  \sigma_z \right] w_y(n \Delta t) \Delta t + \sigma_z  
\eeqn
Each term in summation can be interpreted as a cylindrical Fourier component of the 
x-y magnetization weighted by the RF nutation rate $w_y(n \Delta t)$.  The RF 
waveform specifies the magnitude of each spatial Fourier component, and the 
resulting spatial profile is the Fourier transform of the RF waveform\cite{FourierPicture}.
An equivalent description is to say that, for weak RF pulses, the excited magnetization
of the spins at a given resonance frequency is, to first order, proportional to 
the Fourier component of the RF waveform at that frequency.
As a result, control of the appropriate RF Fourier component 
essentially translates to selective addressing of spatial frequencies, which 
in turn allows the excitation of particular spatial locations.

The Fourier transform approximation allows encoding of arbitrary shapes on 
the various spatial locations of one uncoupled nuclear species.  For QIP, 
however, coupled spins are required to implement two-spin operations.
In particular, the chloroform carbons and protons are coupled together
via the scalar coupling.  Given that the required RF waveforms
should be weak, the coupling interferes with the desired 
evolution.  The effect of the coupling present while encoding on spin 1
is removed by applying a strong RF decoupling sequence on the
second spin \cite{decoupling}.  The decoupling modulates
the $\sigma_z^2$ operator in the interaction Hamiltonian, making its
average over a cycle period equal to zero.  As a result, the second
spin feels an identity operation during the decoupling.  Fig. 2 
shows the complete RF and gradient pulse sequence.  
As can be seen from the diagram, the first encoding
on qubit 1 was subsequently swapped to qubit 2, followed 
by a re-encoding of qubit 1.  We chose this method because the smaller 
gyromagnetic ratio of $^{13}$C causes a narrower frequency 
dispersion in the presence of the gradients, making the carbon 
decoupling simpler.

As described above, the encoding process writes the desired shapes in 
the spatial dependence of each spin's x-magnetization.  The occupation numbers, 
however, are proportional to the z-magnetization, as can be seen
when the number operator in the equation
\begin{equation}
f_a(n, m) = \langle\psi(n,m)|{\hat{n}_a}|\psi(n,m)\rangle,
\end{equation}
is replaced with $\hat{n}_a = \frac{1}{2} (1+\sigma_z^a)$ resulting in
\begin{equation} \label{OccMag}
f_a(n, m) = \frac{1}{2} \left[1 + \langle\psi(n,m)|{\sigma^a_z}|\psi(n,m)\rangle \right].
\end{equation}
where second term in the brackets represents the z-magnetization.  The 
encoding process is followed by a $\pi/2$ pulse that
rotates the excited x-magnetization to the z direction.  

\subsection{Collision and Swap Gates}
After initialization, the next step is to apply the collision operator.
For the QLG algorithm solution to the diffusion equation, 
the collision operator $\hat{C}$ is the square-root of swap gate.  
Expressed in terms of the Pauli operators, it is  
\begin{equation} \label{ImpCollision}
\hat{C} = \text{exp} \left[ -i\frac{\pi}{8} \left( \sigma^1_x \sigma^2_x + \sigma^1_y \sigma^2_y + \sigma^1_z \sigma^2_z \right) \right]
\end{equation}
where an irrelevant global phase has been ignored.  Written in this form, the operation 
$\hat{C}$ can be decomposed into a sequence of implementable RF pulses 
and scalar coupling evolutions\cite{NMRQIPBasics1, NMRQIPBasics2} by noticing that the product operators in the 
exponent commute with each other, resulting in 
\beqn
\hat{C} = \text{exp} \left[ -i\frac{\pi}{8}  \sigma^1_y \sigma^2_y \right] \\
\text{exp} \left[ -i\frac{\pi}{8}  \sigma^1_z \sigma^2_z  \right] \\
\text{exp} \left[ -i\frac{\pi}{8}  \sigma^1_x \sigma^2_x  \right]
\eeqn
Expanding the first and last exponentials as scalar couplings
sandwiched by the appropriate single-spin rotations results in
\beqnar \label{finalCol}
\hat{C}&=& \text{exp} \left[  i\frac{\pi}{4} \sigma^1_x \right] \text{exp} \left[  i\frac{\pi}{4} \sigma^2_x \right] \text{exp} \left[ -i\frac{\pi}{8}  \sigma^1_z \sigma^2_z  \right] \text{exp} \left[ -i\frac{\pi}{4} \sigma^1_x \right] \text{exp} \left[ -i\frac{\pi}{4} \sigma^2_x \right] \cdot \nonumber \\
& & \text{exp} \left[ -i\frac{\pi}{8}  \sigma^1_z \sigma^2_z  \right] \cdot  \\
& & \text{exp} \left[ - i\frac{\pi}{4} \sigma^1_y \right] \text{exp} \left[ - i\frac{\pi}{4} \sigma^2_y \right] \text{exp} \left[ -i\frac{\pi}{8}  \sigma^1_z \sigma^2_z  \right] \text{exp} \left[ i\frac{\pi}{4} \sigma^1_y \right] \text{exp} \left[ i\frac{\pi}{4} \sigma^2_y \right] \nonumber
\eeqnar
The exponents of terms proportional to $\sigma^1_z \sigma^2_z$ represent internal
Hamiltonian evolutions lasting for a time  $t_{zz}^{col} = 1/(4 J)$.  The 
exponents of terms with single-spin operators are implemented by $\pi/2$ 
rotations.  They were generated by RF pulses whose nutation 
rate was about 50 times greater than $J$.  All of the pulses and delays were 
applied without a magnetic field gradient in order to transform all of the sites 
identically.

As shown in Fig. 2, swap gates were utilized both in the lattice initialization
and in the measurement of the carbon magnetization.  The pulse sequence for the swap 
gates was almost identical to the sequence for $\hat{C}$.  The only difference was
that the internal evolution delay was set to $t_{zz}^{swap} = 1/(2 J)$.

\subsection{Measurement}
The occupation numbers resulting from the collision were obtained 
by measuring the $z$-magnetizations and using equation (\ref{OccMag}).
Since only the $\sigma^a_x$ and $\sigma^a_y$ operators
are directly observable, a ``read out'' $\pi/2$ pulse  
transformed the $z$-magnetization into $x$-magnetization.  The proton 
magnetization was measured directly 
after the collision, while the carbon magnetization was first swapped to
the protons before observation.  Measurements of both the $^{13}$C and $^{1}$H 
magnetizations were carried out separately, and in both cases via
the more sensitive proton channel.  The measurements were made in the
presence of a weak linear magnetic field gradient, causing signals 
from different sites to resonate with distinguishable frequencies.
The observed proton signal was digitized and Fourier transformed to 
record an image of the spatial variation of the spin magnetization.
The observed spectrum was then processed to correct the baseline 
and to obtain the resulting magnetization at each site.  Because each site 
is composed of a slice of the sample with spins resonating in a band of 
frequencies, the occupation number for each site was obtained by 
averaging over all spins in the corresponding band.

\subsection{Streaming}
The final step involves classically streaming the results of the
measurements according to equations (\ref{streamOcc1}) and (\ref{streamOcc2}).  
The streaming operation is applied in conjunction with the next lattice 
initialization step by adding a linearly varying phase to the Fourier 
transform of the desired shape.  The added phase causes a shift in the 
frequency of the pulse determined by the slope of the phase.
When the frequency-shifted pulse is applied in the presence of the 
magnetic field gradient, the shift results in spatial translation 
of the encoded shape.  The streaming operation is thus implemented as a 
signal processing step in the lattice initialization procedure.

\section{Results and Discussion}

The results of the experiment are shown in Fig. 3, together with plots
of the analytical solution and of numerical simulations of the NMR experiment.
In total, 7 steps of the algorithm were completed using a parallel array of 16 
two-qubit ensemble NMR quantum processors.  The observed deviations
between the data points and the analytical plots can be attributed 
to imperfections in the various parts of the NMR implementation.  

To explore the source and relative size of these errors, we 
simulated perfect experiments, each time adding controlled 
errors in four sections of the implementation:
\begin{itemize}
\item	Fourier transform approximation in the initialization
\item	Decoupling during the initialization
\item 	Encoding swap gate and $\pi/2$ pulse errors
\item	Collision gate errors
\end{itemize}
The Fourier transform approximation executes a correct writing
of the desired magnetization to first order in the overall
flip angle.  To explore errors introduced by the approximation,
we simulated NMR experiments using nutation angles ranging
from $\pi/2$ to $\pi/20$.  In this range, angles smaller than $\pi/4$ 
resulted in accurate encodings of the desired Gaussian shapes
through the ten steps of the implementation.  The errors 
in the three remaining sections were simulated by using RF pulses 
with the actual time and nutation rate that were used on 
the spectrometer. By using a finite power, errors from imperfect
averaging of the scalar coupling could be observed.  Errors 
in the collision gate caused the least impact to the mass density,  
followed by errors originating from the imperfect decoupling 
sequence.  The largest deviations originated from realistic 
simulations of the swap gate and the $\pi/2$ pulses in 
the encoding.  It is important to note that the simulated 
gate fidelities for the swap and collision gates, although imperfect, 
are still about $0.995$.  This suggests that the observed deviations are
caused by the coherent buildup of errors through a few iterations,
and not just by the individual errors from a single gate.
The complete simulation, using realistic RF
pulses and a shaped pulse nutation angle of $\pi/4$,
is plotted in Fig. 3.  The calculated mass densities closely match
the experimental results, suggesting that the observed errors 
are accurately modeled.  

Other potential sources of errors include the finite signal to noise, the
state fidelity of the starting pseudo pure state, and gradient
switching time.  In addition, spin relaxation, random self-diffusion 
of the liquid molecules, and RF inhomogeneity can all cause attenuations
in the strength of the signal.  In our experiments, these last three
mechanisms manifested themselves indirectly through reduced signal to noise.
However, given that this attenuation was present in all of the experiments,
any direct results were mostly normalized away in the data processing.
Although none of the above errors contributed significantly to our 
implementation, they are likely to become important as more complicated 
algorithms are executed on larger lattices.  

In particular, molecular diffusion over the time of an operation places a lower 
bound on the physical size of the volume element corresponding to each site in 
the computation.  In the 1-D case discussed here, the root-mean-squared displacement
($\Delta z= \sqrt{2Dt}$) for chloroform ($D=2.35\times10^{-5} cm^{2}/s$) is about 
$10.8 \mu m$ over the $25ms$ needed for encoding and the collision operator.  Since 
the actual volume element were about $625\mu m$ across, this resulted in a negligible 
mixing of the information in adjacent sites.  
However, it is clear that for this approach to type-II quantum 
computer to remain viable for large matries and more complex collision operators  
the physical size of the sample must grow with the size of the problem.


\section{Conclusion}

Ensemble NMR techniques have been used to study the
experimental details involved in quantum information processing.  
The astronomical number of individual quantum systems ($\sim 10^{18}$)
present in typical liquid-state spin ensembles greatly facilitates
the problem of measuring spin quantum coherences.  In addition, the 
ensemble nature has been successfully utilized to create the necessary
pseudopure states\cite{CoryPP,IkePP} and to systematically generate 
nonunitary operations over the ensemble\cite{HavelGradDiff}.  
In this experiment, we again exploit the ensemble nature, but 
this time as a means of realizing a parallel array of quantum 
information processors.  The novel architecture is then used to run 
a quantum lattice gas algorithm that solves the 1-D diffusion equation.  

The closeness of the data to the analytical results is encouraging, 
and it demonstrates the possibility of combining the advantages of 
quantum computation at each node with massively 
parallel classical computation throughout the lattice.  Currently, 
commercial MRI machines routinely take images with $256\times256\times256$ 
volume elements.  As a result, the large size of the NMR ensemble 
provides, in principle, sufficient room to explore much larger lattices.  
However, in moving to implementations with more computational power, several 
challenges remain.  The limited control employed here is 
sufficient for a few time steps of the algorithm, but refinements
are necessary to increase the number of achievable iterations.  
In addition, although complicated operations have been done in up 
to 7 NMR qubits\cite{SevenQubits1,SevenQubits2,SevenQubits3}, the problem 
of efficiently initializing a large lattice of few-qubit processors still 
remains.  Our results provide a first advance in this direction, and they 
provide confirmation that NMR techniques can be used to test these 
new ideas.

\section{Acknowledgments}
We thank E.M. Fortunato and Y. Liu for valuable discussions.  This 
work was supported by the Air Force Office of Scientific Research.

\newpage
\begin{figure}
     {\centerline{\epsfig{file=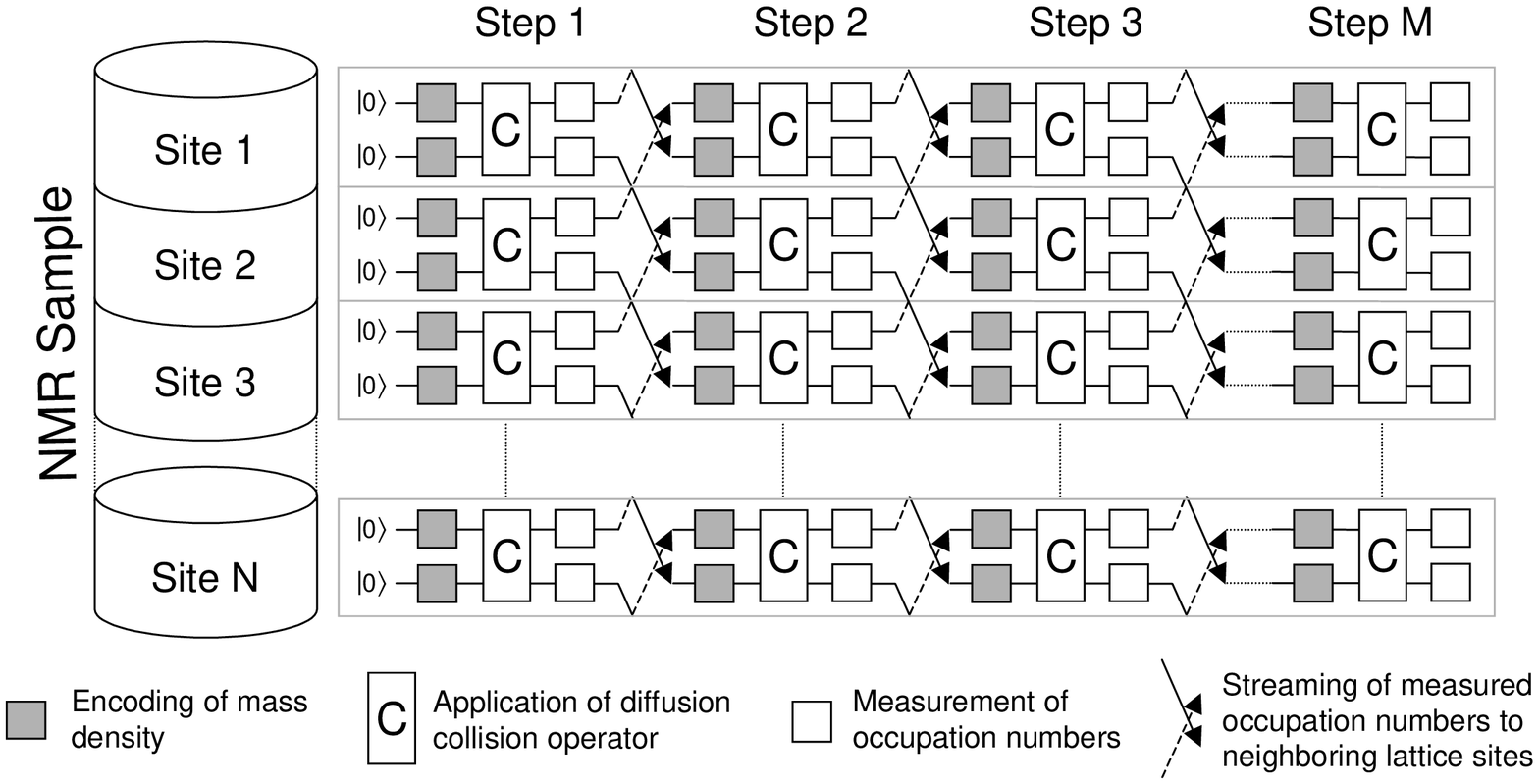, angle=0, width=\linewidth}}}
        \vspace{.15cm}
        \caption{The circuit diagram shows the quantum lattice gas algorithm for solving 
the 1-D diffusion equation.  The algorithm employs $N$ two-qubit sites 
to encode the discretized mass density.  Each site codes for a single
value of the mass density using the quantum state of the two qubits.
The encoded information is subjected to a series of local transformations
that evolve the system.  The collision operator $C$ is the only potentially entangling
operation in the algorithm, and it creates quantum coherences limited to 
each two-qubit system.  The streaming is executed by classical communication, 
and it moves the occupation numbers up and down the lattice as denoted by the 
arrows.  The sectioned cylinder depicts the position of 
the adjacent sites in the NMR sample.  Each site is physically realized as an 
addressable slice of isotopically-labeled Chloroform solution.\vspace{.1in}
}
        \label{algorithm}

\end{figure}
\newpage

\begin{figure}
      {\centerline{\epsfig{file=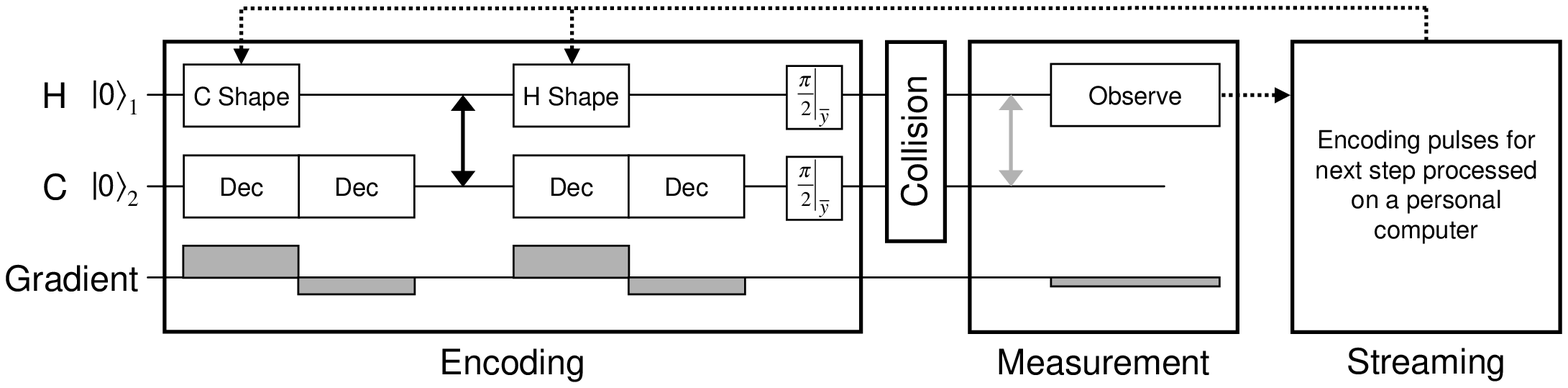,angle=0, width=\linewidth}}}
        \vspace{.20cm}
        \caption{The NMR implementation consists of four main sections, each corresponding 
to the prescribed QLG algorithm step.  The top two lines in the diagram 
correspond to RF pulses applied to the proton and carbon qubits, respectively.
The third line shows the application of magnetic field gradients.  In the
encoding section, the initial carbon magnetization is recorded on the
protons before being transferred to the carbons.  The starting magnetization
is specified by using a RF pulse shaped as the Fourier transform of
the desired magnetization.  The shaped pulses are applied in the presence 
of gradients so that each site can be addressed.  A carbon decoupling sequence
prevents the scalar coupling from interfering with the low power shaped pulses.
The $\pi/2$ at the end of the encoding move the information form the x-axis
to the z-axis, as required by the QLG algorithm.  The collision operator follows the
encoding, and it is implemented without gradients to ensure that all of the
sites in the sample feel the same transformation.  The results are observed
in two experiments, each time using the more sensitive proton channel.  A
swap gate is added when measuring the carbon magnetization.  Finally, the
streaming operation is applied by shifting the frequencies of the carbon
and proton shapes in opposite directions.  }
        \label{pulseSeq}

\end{figure}

\newpage
\begin{figure}
       {\centerline{\epsfig{file=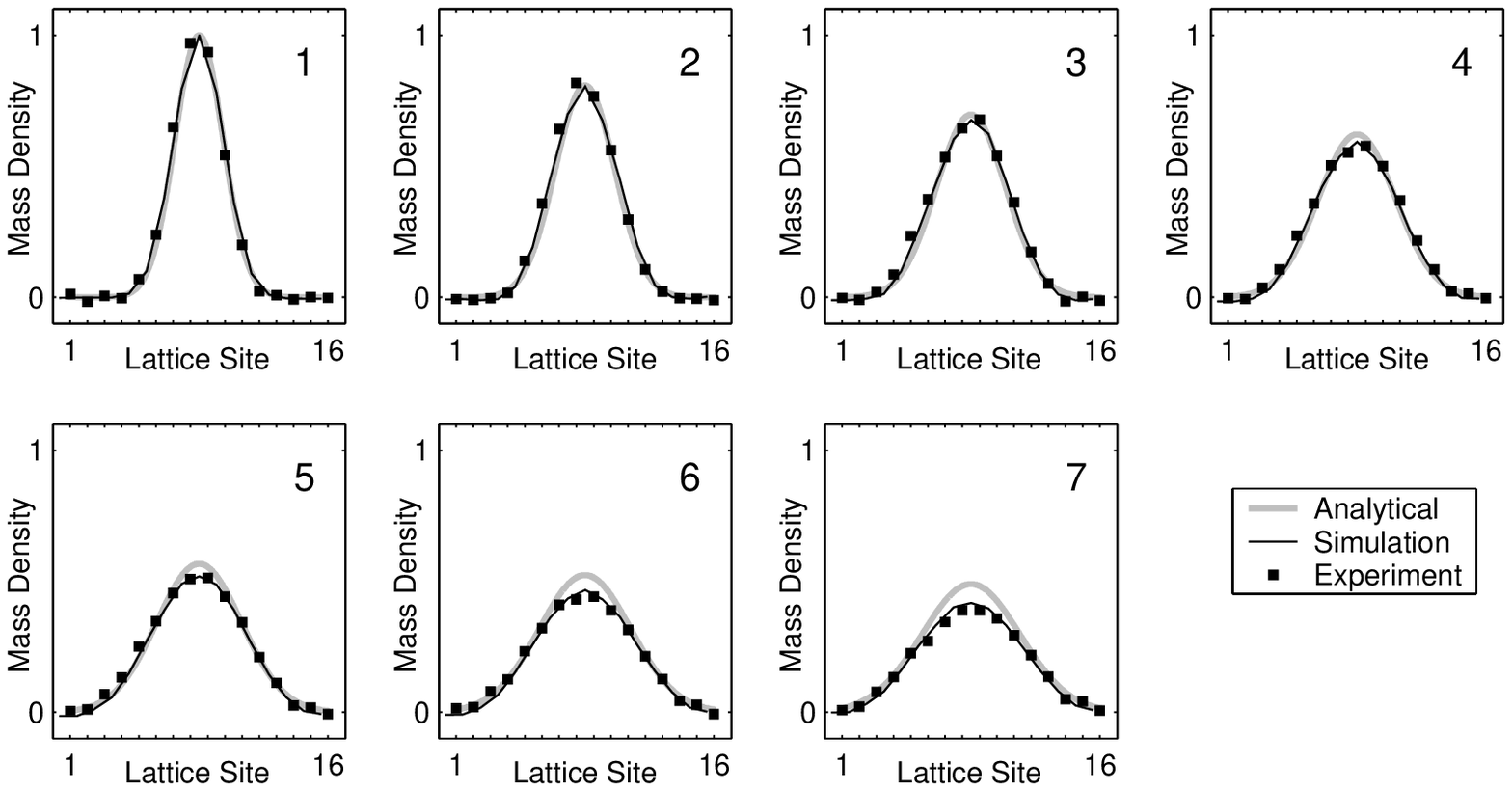, angle=0, width=\linewidth}}}
        \vspace{.20cm}
        \caption{The experimental mass densities are plotted in the figure, 
together with plots of the analytical solution and the numerical simulation of 
the NMR experiment.  The normalized, dimensionless mass densities are plotted
as they were encoded on the lattice.  Seven steps of the algorithm were 
implemented on 16 two-qubit sites.  The simulations 
were performed using the actual RF nutation rates and times of the
experimental setup.  The calculations closely match the data, suggesting 
that the deviation between the analytical results and the data can be attributed
imperfections in the methodology.  As a result, the simulations promise to be 
useful in exploring the errors from alternate methods. }
        \label{data}
\end{figure}


\begin{thebibliography}{99}
\bibitem{decoupling}
	The decoupling was accomplished by applying 
	the pulse cycle $Q \bar{Q} \bar{Q} Q$ during the positive and negative 
	gradients.  The element $Q$ is a composite $\pi$ pulse implemented
	with four sequential pulses having nutation angles
	$80.4^\circ, 362.0^\circ, 181.6^\circ, 180.8^\circ$ and
	respective phases $271.3^\circ, 132.4^\circ, 292.3^\circ, 200.4^\circ$.
	This composite pulse was chosen over more commonly used pulse
	sequences for its relatively short total nutation angle 
	and good decoupling, allowing the cycle to fit within a 
	gradient period.
\end{thebibliography}
\end{document}